\documentstyle[prd,preprint,tighten,aps,preprint]{revtex}
\begin{document}
\draft
\preprint{
\begin{tabular}{r}
JHU-TIPAC 940020
\\
DFF-213/11/94
\end{tabular}
}
\title{Do the Age of the Universe and the Hubble Constant Depend on What Scale
One Observes Them ?}
\author{C. W. Kim\thanks{E-mail Address, CWKIM@JHUVMS.HCF.JHU.EDU}}
\address{Department of Physics and Astronomy \\
The Johns Hopkins University\\
Baltimore,Maryland 21218\thanks{Permanent Address}\\
and \\
Dipartimento di  Fisica, Univ. di Firenze\\
I.N.F.N. Sezione di Firenze, Firenze, Italy}
\maketitle

\begin{abstract}
\setlength{\baselineskip}{.5cm}
The apparent cosmological conflict between the age of the
Universe, predicted  in the standard Friedman cosmology by using the recent
measurement of the larger Hubble constant from a direct calibration of
the distance to the Virgo galaxy cluster,
and the ages of the oldest stars and globular clusters is resolved by
invoking the scale dependence of cosmological quantities, including
the age of the Universe. The distance dependence or the running of
cosmological quantities is motivated by the asymptotically-free higher-
derivative quantum
gravity. The running can also be derived by ``properly" modifying the Friedman
 equations. This property
can also provide partial explanation of the apparent disagreement between
the two
recent measurements of the Hubble constant using NGC 4571 at 15 Mpc and
NGC 5253 at 4 Mpc.
\end{abstract}


        Recent measurements of the Hubble constant, $H_0$   using NGC 4571
(distance =$14.9 \pm 1.2$ Mpc) in Virgo cluster \cite{virgo1} and NGC 5253
(distance = 4.1 Mpc) in a galaxy closer than Virgo cluster \cite{virgo2}, and
their  implied ages of the Universe have become a subject of heated
controversy. In addition to the disagreement between the two measurements,
the major problem is that the age of the  Universe  predicted by the use of
the larger value of $H_0$  from  NGC 4571 ($H_0 = 87 \pm 7$ Km/sec Mpc)
and the standard Friedman cosmology
with flat space becomes about half the measured ages of 14 to 18 Gyr
for the oldest stars and globular clusters \cite{star}.
In this article, we show that the application of the concept of the  running of
the physical quantities according to the Renormalization Group Equations (RGE)
\cite{RGE} to cosmology can solve the age puzzle and explain the difference
 between the two $H_0$ measurements, at least partially. It is
well-known that masses as well as coupling constants
are running, i.e. their values depend on what energy or momentum
scales one measures or calculates them. If the gravitational constant $G$ is
asymptotically free as suggested by recent works \cite{gcG}, ``modified"
Friedman-like cosmological equations are shown to imply that $G$, $ H_0$,
and $\Omega_{0}(=(\rho_{0}/\rho_{C}))$ are all increasing functions
of scale or distance, even at present epoch, whereas the age of the Universe,
$t_0$, is a decreasing function of distance at which one measures or calculates
it.

There are several ways to modify the Friedman cosmology. The best is, of
course, to start with ``the ultimate theory" of quantum gravity. Since this
has not been
realized as yet, one can approach in the way described in Refs.\cite{five}
and \cite{six} by incorporating the running gravitational constant into the
Friedman equations. One can also modify the Robertson-Walker metric
by relaxing the Cosmological Principle, which states that the Universe has
symmetric homogeneous space-time, with  "homogeneous" replaced by "only
globally homogeneous". This leads to the running of $H_{0}$, $R_{0}$ and
$\Omega_{0}$ as functions of scale. To accomplish this, it is also
necessary to modify the Einstein equation itself by replacing its right-hand
side by $8 \pi G(d )T_{\mu\nu}$ with the constraint that the
covariant derivative of $G(d)T_{\mu\nu}$ be zero where $d$ denotes
 distance. This, of course, modifies
the equation of state and mimics the running gravitational constant as a
function of distance. Details on this approach will be discussed
elsewhere \cite{seven}.

        To set the stage for later cosmological argument, let us consider
the following
metric in order to illustrate the case of simply changing the metric,
\begin{equation} d\tau^2=dt^2-R^2(t,r)(dr^2+r^2d\Omega^2).
\end{equation}
When $R( t, r ) =R( t )$, Eq.(1) reduces to the Robertson-Walker metric for
$k$ = 0, which is a consequence of the Cosmological Principle. Equation (1)
describes the Universe which has spherically symmetric ( and only globally
homogeneous) space-time. Physical rational for this relaxing is as follows.
When one looks at a large scale object such as, say, a cluster, then due
to the increasing amount of dark matter and/or larger gravitational
constant, it appears that $\Omega_{0}$ has increased from that of a
galaxy scale. Since the scale factor is governed by $(G\rho)$ which now
depends on scale, it must depend on the scale $r$ as well as $t$ ( see
 \cite{seven} for detail).

For an illustrative purpose, let us consider the case when
\begin{equation}
R(t,r)= a(t)\left[ 1 + \alpha \left (ra(t)\right )^{n}\right] ;
\alpha > 0, n> 0,
\end{equation}
which is clearly an increasing function of $r$. That is, the local $(r=0)$
scale factor at any given time is always the smallest, and the scale
factor at any fixed time increases as distance ( from us ) increases.
In this case  the metric is obviously not that of the Robertson-Walker.
(The case, in which $R(t,r)$ is factored out as $a(t)S(r)$, reduces to the
Robertson-Walker metric with  redefinition of $r$.)
The the expansion rate becomes
\begin{equation}
H(t,r) = \frac  {\dot R}{R}=
 \left[ \frac {\dot a}{a}\right]  \frac {
1+(n+1)\alpha r^{n}\left[ a(t)\right] ^{n}} {1+\alpha r^{n}\left
[ a(t)\right]^{n}}\;\; ,  \end{equation}
which is now an increasing function of $r$. Stricktly speaking, the
expansion rate must be defined by $({\dot D}/D)$, where $D(t,r)=
\int R(t,r)dr$, but
for scales up to, say, the Virgo cluster which  has
$z \sim 0.004$ and $r \sim 4 \times 10^{-3}$ (assuming
the horizon has $r \sim 1$), the definition used in Eq.(3) is , for our
present purpose, a good approximation to the more rigorous one.
In general,
however, one needs caution in defining the expansion rate in the modified
Friedman cosmology, especially for dealing with much more distant or larger
objects.

The present value, $H_{0}(r)$, in Eq.(3), is no
longer universal, but instead depends on what scale one uses to observe
or calculate  $H_0$. If $H_{0}^{2}(r)$ is proportional to the gravitational
constant $G$ as in the  case of the Friedman cosmology,  $G$
increases as $r$ increases as well.  Also, since the age of the Universe,
when it is calculated
using the  cosmological equations, is inversely proportional to
$H_{0}(r)$,
the age of the Universe becomes older when estimated using a smaller
scale, whereas it decreases as the scale becomes larger.
This does not mean, however, that the age
depends on where one calculates. In fact, every observer who uses
the same scale $r$ at $t_0$ would obtain the same age. Although the above
discussion has nothing to do with the RGE, physics involved is precisely
that of the running of masses and coupling constants as functions
of energy or momentum scale based on the RGE. As in the case of
running masses and coupling constants, direct comparison of cosmological
quantities makes sense only when they
were observed or calculated at the same scale. The same quantity can take
different values at different scales. Therefore, the measured ages of
the oldest stars and globular clusters in our neighborhood (less
than 100 Kpc, say) have to be compared with the age of the Universe
calculated at the same distance scale. Now the question is `` How can this
scenario be realized ? "  Here, we follow the approach used in Refs.
\cite{five} and \cite{six}.

In the asymptotically-free higher-derivative quantum gravity \cite{gcG},
an inflationary period eventually settles into a standard Friedman epoch that
behaves as if the Universe is matter-dominated, but with a very important
modification. That is , the gravitational {\it constant} that appears in the
Friedman equations is replaced by the one which is asymptotically free,
i.e., $G$ increases as a function of distance scale according to the
RGE \cite{gcG,five}. This $G$
takes the value of the Newton's value $G_N$  at short distances
but slowly starts to rise as distance increases.
For this asymptotically- free behavior, see, for example, figure 2 of
Ref.\cite{five} and figure 1 of \cite{six}. In addition to the above
modification,  there are other consequences.

During the inflation, space is flattened so that the $k = 0$ case (flat)
is realized. Therefore, the present value of $\Omega_{0}\equiv
(\rho_{0}/\rho_{c})$ is unity, where $\rho_{0}$ is the present density
 and $\rho_{c}$ is the present density in the case of a flat
Universe. However,
there is a very important difference between the standard cosmology and
the present one. Since $G$ is distance-dependent ( or running), $\Omega_0$
also becomes
distance- dependent. That is, $\Omega_0$ does {\it not} have to be unity
everywhere. $\Omega_0$ = 1 may be realized at a very large distance, maybe
near the horizon or at some point further out than the horizon. Therefore,
in our local neighborhood $\Omega_0$ is less than unity and locally open even
though the Universe as a whole has no curvature. (This point has also been
noticed by others in different context \cite{eight}.) In short, some
cosmological quantities may run as functions of distance scale.
The running of $H_0$ and $\Omega_0$ and others
can, in principle, be calculated based on their RGE, but unfortunately
it is not feasible at present because of the lack of a satisfactory
quantum version of gravity . Therefore, we adapt the following
phenomenological approach.

The modified Friedman-like equations with $ k = - 1$
( this is true in our local neighborhood
since the local $\Omega_0$ is less than unity), that
have been suggested by the asymptotically-free higher-derivative quantum
gravity, are \cite{five,six}
\begin{equation}
{\dot R}^2(t,r)={\frac{8\pi}{3}}\rho G(d)R^2(t,r) + 1 \; ;\;\; G\cdot
\rho = G\cdot\rho_0
\left[ {\frac {R_0}{R}}\right]^3 \; ,
\end{equation}
\begin{equation}
{\Omega}_0(d_0)= \frac {G\cdot \rho_{0}}{G\cdot \rho_{c,0}} =
\frac {8 \pi G(d_0)\rho_0}{3{H_0}^2(d_0)},
\end{equation}
where the subscript zero denotes the present values, and
 $G(d)$ is the value of $G$ at (proper) distance $d$
, if one interprets $d$ as an inverse of momentum, or simply $r$ if one
uses the  comoving distance parameter.  But, at present epoch, this
distinction is irrelevant. The validity of
the above equations is, admittedly, not on a firm ground as one would
like it to be. But, in this article, we shall assume them as
phenomenological and treat them as such. With this cautionary remark,
we now investigate some consequences of Eqs.(4) and (5). The behavior of
$G(d)$ given in Ref.\cite{five} and \cite{six} can be best fitted by the
following  expression:
\begin{equation}
G(d)= G_N (1+ 0.3 \; d\, ^{0.15}) \equiv G_N \left[ 1+ \delta_G(d) \right],
\end{equation}
where $d$ is expressed in units of Kpc \cite{six,seven}. Since $d$ is
 a proper distance, we can rewrite
\begin{equation}
d(t,r)= \left[ \frac {R(t,r)}{R(t_0,r)} \frac {{10^7}r}{3} \right],
\end{equation}
where $r=1$ corresponds roughly to the size of the horizon and
$r =3\cdot 10^{-7}$
gives 1 Kpc at present. In fact,  in Eq.(7) $d$ does not have to depend
on $t$ as mentioned already. Such a model is presented in Ref.\cite
{seven}.  It is to be emphasized here that an analytic solution for
 $R(t, r)$ cannot be obtained from Eq.(4) but  since $d$ is
a function of $t$ and $r$, $R$ obviously has to be a function of $t$ and $r$.
This behavior
is precisely the same as that of the example discussed in Eq.(2). As a
consequence,  the expansion rate $H = ({\dot R}(t,r)/R(t,r))$
( recall that this approximation is valid for $z\ll 1$) is also
a function of $r$
or distance. As mentioned already, $\Omega_0$ also depends on the
distance scale.
That is, both $H_{0}$ and $\Omega_0$ run as functions of distance.

An interpretation of Eq.(5) as an indication of growing $\Omega_0(d_0)$
with distance was discussed in Ref.\cite{five}, leading to the necessity of
less or no dark matter in the Universe. In arriving at this conclusion, it was
assumed in Ref.\cite{five} that $ H_0$  has no distance dependence. It was also
shown in Ref.\cite{six} that Eq.(7) leads to an increased power of the two
point  correlation function at large distances without help of dark matter
with        $\Omega_0= 1$. In Refs.\cite{six} and \cite{eight}, the possibility
of {\it growing $H_0$ with scale} was
also mentioned with no further elaboration.

Equations (4) and (5) lead to the following
phenomenological expressions for the running behavior of $H_0$ and $\Omega_0$:
\begin{equation}
H_{0}(d_0)\simeq  {\overline{H}_{0}}\sqrt { 1+ {\overline{\Omega}_0}
\delta_G(d_0)},
\end{equation}
\begin{equation}
\Omega_0(d_0)\simeq {\overline{\Omega}_0}\frac{1+\delta_G(d_0)}
{1+{\overline{ \Omega}_0}
\delta_G(d_0)},
\end{equation}
where bars denote local values at, say, $r\simeq 0$ or $d_{0} \simeq 0$ and
$\delta_G(d_0)$ is given by Eq.(6). Both $H_0(d_0)$ and $\Omega_0(d_0)$
are increasing  functions of $d_0$. However,the scale dependence of $H_0(d_0)$
is much slower than that of $\Omega_0(d_0)$. In deriving Eqs.(8) and (9),
we have assumed that the $d_0$ dependence of $H_0^2$
is mainly due to that of $G$ so that the $d_0$ dependence of $R(t_0,d_0)$
is, in general, much slower than those of $H_0(d_0)$ and $G(d_0)$ (i.e.,
$R(t_0,d_0) \simeq R(t_0,d_{0}\simeq 0)$ for up to the distance of our
interest, 15 Mpc.
This assumption is explicitly justified in a model considered in Ref.[8].
 We must also caution that Eqs.(8) and
(9) did not take into account a further increase of $\Omega_0$ due to the
possibility that the dark matter  content increases as distance increases.
(This
effectively makes $\delta_G(d_0)$ grows faster than Eq.(6) if one keeps
$\rho_0$
as constant.) Equation (8) states that $ H_0$  obtained
at distance of 15 Mpc has to be somewhat larger than that of 4 Mpc since
$H_0$ is slowly
increasing  as distance increases, which explains
the discrepancy mentioned above, at least qualitatively. Our
estimate based on Eq.(8) without dark matter indicates that
the growth of $G(d)$ is not sufficiently fast enough to explain the
discrepancy between the recent two measurements of $H_0$, although up to about
20\% increase of $H_0$ from 4 Mpc to 15 Mpc can be accounted for. However,
within one and half standard deviations, the two measurements can be  made
consistent in our picture of growing $H_0$ without dark matter.
The increase of $\Omega_{0}(d_0)$ is shown to be consistent with observations
 (see also Ref.[6]) with much less dark matter than $\Omega_{0} = 1$ requires.
 Results of detailed calculations with (and without) the contributions
 from dark matter  will be given elsewhere \cite{seven}.

We now proceed to discuss the age problem. Combination of Eqs. (4)
and (5) yields, with the approximation mentioned already,
\begin{equation}
\frac {1}{{R_0}^2(d_0)}\simeq \frac{1}{{R_0}^2(d_0 \simeq 0)} \simeq
 \left[ 1-\Omega_0(d_0)\right] {H_0}^2(d_0).
\end{equation}
Note that in our approximation the distance dependences
of $\Omega_0(d_0)$ and $H_0(d_0)$ on the right- hand side are supposed to
be cancelled with each other.

Using Eqs. (4), (8), (9), and (10), we find the age equation
\begin{equation}
\int^{1}_{0} \frac  {\sqrt{x} dx}{\sqrt{x\left[ 1-\overline{\Omega}_0(d_0)
\right] +
{\overline{\Omega}}_0(d_0)\left[  1+\delta_G(x,r)\right]}}
\simeq \overline{H}_{0} t_0(d_0),
\end{equation}
where
\begin{equation}
\delta_G(x,r) = 0.3\left[ \frac {10^{7}rx}{3} \right] ^{0.15},
\end{equation}
and
\begin{equation}
x = \frac {R(t,r)}{R(t_0,r)}.
\end{equation}
An identical expression to Eq.(11), with exception that
$\delta_{G}$ has no $x$ dependence,
can also be derived  classically  by modifying the Robertson-Walker metric
and the Einstein equation \cite{seven}.
The following comments are in order:
\begin{enumerate}
\item When $\overline{\Omega}_0$ and $\overline{H}_0$ are replaced
by the standard $\Omega_0$
    and $H_0$, and $\delta_G(x,r)$
    is set equal to 0, Eq.(11) reduces to the age equation
    for $k = - 1$  in the Friedman cosmology.
\item Equation (11) is valid up to the scale where our approximate
    definition of $H$  is still valid.
\item The age of the Universe $t_0(d_0)$ depends on the value of $d_0$ or
    $r$, i.e.,
    $t_0(d_0)$ is also running as a function of $d_0$ so that the age
    depends on which distance or scale we are interested in.
    As can be seen from Eqs.(11) and (12), the smaller the value of $r$,
    the older the age becomes and vice versa. This is our main result.
\end{enumerate}

We now present some results of numerical calculations based
on Eq.(11). Interestingly, it turns out that as long as $d_0
\lesssim$100 Kpc, $t_0(d_0)$ in our neighborhood
is insensitive to
 $d_0$. It only depends mildly on the choice of $\Omega_0(d_0)$.
We find, for $d_0 \lesssim$ 100 Kpc,
\begin{eqnarray}
{\overline{t}}_0 \simeq \frac {0.9}{{\overline{H}}_0}\; \;
\text{for}~~ {\overline{\Omega}}_0 = 0.1 \nonumber\\
{\overline{t}}_0 \simeq \frac {0.85}{{\overline{H}}_0}\; \;
\text{for}~~ {\overline{\Omega}}_0 = 0.2,
\end{eqnarray}
where ${\overline{H}}_0$ and ${\overline{\Omega}}_0$ denote our
"local" values, as defined earlier.
We believe that ${\overline{\Omega}}_0 =
0.1$ case is more appropriate because this value is consistent
with the dark matter content in our galactic halo as well as with the
bound from nucleosynthesis.
The choice of the local value of ${\overline{H}}_0$ is
 somewhat ambiguous due to
our inability to measure it locally. Therefore, we take, for definiteness,
the most commonly used canonical lower value ${\overline{H}}_0 =
50$ Km/sec Mpc.
The resulting age is
\begin{equation}
{\overline{t}}_0 \simeq 18 \; \text{ Gyr },
 \end{equation}
which is comfortably long enough to accommodate the known ages of oldest
stars and globular clusters , also obtained in our neighborhood. We emphasize
that this comparison is justified because both are measured or calculated
at the same scale.

The age at the distance of 15 Mpc becomes
\begin{equation}
t_0(d_0 = 15 Mpc) \simeq \frac {0.81}{\overline {H}_0} = 16  \; \text{Gyr} ,
\end{equation}
which is shorter than that of Eq.(15). This, however, is
anticipated , as mentioned already, because of the running (decreasing as
scale increases) of $t_0(d_0)$.

Another age of interest is the one at the distance of 4 Mpc.
The result  is $t_0(d_0) \simeq 16.4$ Gyr.
We mention here that because of the approximation we used in defining
$H$, we cannot use Eq.(11) to calculate the age at scales much beyond
the Virgo cluster.
For comparison, the standard
Friedman cosmology with $\Omega_0 =1$ gives the age
\begin{equation}
t_0 = \frac {0.66}{H_0},
\end{equation}
where $H_0$ is a fixed value, independent of scale. Therefore, $t_0$ in Eq.(17)
does not run. This causes problems when different  observed values of $H_0$
are used in Eq.(17). It is worth mentioning that
in order to be self-consistent, one has
to take into account the effective growth of $\delta_G(x,r)$ in Eq.(11) due to
dark matter in addition to that given in Eq. (6). This is because we
have used {\it effective} $\Omega_{0}(d_0)$ values such as
0.1 for $d_0 \lesssim$
 1 Kpc
and 0.4 for $d_0=15$ Mpc
which include the possible dark matter contributions.
In order to see the sensitivity of the age on the content of dark matter,
we have increased
the coefficient 0.3 in Eq. (12) by 50 \%  which requires a large amount
of dark matter. The resulting ages turn out to decrease
only by 2 \% and 8 \%, respectively, for $d_0 \lesssim$ 100 Kpc and $d_0 = 15$
Mpc.

The final, but important comment on the age calculation concerns
the interpretation of various running ages. First, as mentioned already,
direct measurements of the ages of local stars and globular clusters do not
depend on the use of $H_0$.
Therefore, their ages should be compared with the age of the Universe
estimated at the same scale, i.e., with the so-called local age of the
Universe given in Eq. (15). The agreement is excellent. Then, what does
the age of, for example, 16 Gyr
at the distance of 15 Mpc mean ?   It is simply the age for that scale
which cannot be compared directly with the ages obtained at different scales,
as the fine structure constant $\alpha$ measured at low energies is
different from
the one measured at LEP energy.

In summary, we have demonstrated that in a  modified Friedman-
like cosmology based on the asymptotically-free quantum gravity, admittedly
phenomenological in nature, even the
present values of $\Omega_0$, $H_0$ and $R_0$  all increase as the scale
 increases. We can
make the two recently measured values of $H_0$ consistent with each other
with this growing $H_0$ (with scale) within one and half standard deviations
, only because $H_0$ does not increase fast enough.
We have also shown that  when one calculates the age of the Universe,
the result
depends on what scale one uses. In fact, the age, $t_0$, is a
decreasing function of scale. The
calculated age of the Universe in our  local neighborhood is about 18 Gyr
for the assumed local values ${\bar\Omega}_0= 0.1$ and  ${\bar H}_0 = 50$
Km/sec Mpc, consistent with the ages of the oldest stars and globular
clusters, 14 to 18 Gyr. Our neighborhood is locally open with
${\bar \Omega}_0 = 0.1$ ,
although the Universe as a whole is flat.

In contrast, the calculated ages of the Universe at distances of
4 Mpc and 15 Mpc from us
 are 16.4 and 16 Gyr, respectively. We
argue that these ages  are simply not our local ages and thus
should not, in principle,
be compared with the ages of the oldest stars or globular clusters
in our neighborhood, although they agree within errors.
Instead, they should be used in relating quantities
observed or calculated at the same scales.

How does the running of $G$ affect the early Universe? Let us take the
nucleosynthesis as an example. The relevant $G$ calculated at the horizon
scale within which microphysics operates at the time of nucleosynthesis
is the same as  $G_{N}$  since $d$ in Eq.(7) and thus $\delta_{G}(d)$
in Eq.(6) are practically zero.
Therefore, the nucleosynthesis proceeds in the standard manner.

It goes without saying that it is important to continue further
study of gravity beyond that of Einstein's because cosmological implications
are far- reaching. Only when we have a satisfactory understanding of quantum
gravity on hand, the validity of crucial equations such as those  in Eqs.(4)
and (5) can be critically checked and the RGE's for various cosmological
quantities can directly be derived. An entirely different, but classical
approach based on the modification of the Robertson-Walker metric by
relaxing the Cosmological Principle and the Einstein equation so that
$G$ runs with scale, and further
details on how $H_{0}(d)$ and $\Omega _0(d)$ behave as
functions  of scale, as well as  their cosmological
implications are  given elsewhere \cite{seven}.

We close with the following remark. In our modified cosmology,
the standard formulas that relate distance, redshift, $H_0$ and so on
are modified to accommodate the scale dependence of $R(t,r)$.
Although our results are valid up to the scales where the approximate
definition of $H$ is still valid (i.e., for small $z$), the expected
modifications could be
significant for larger redshifts when
the correct definition of $H$ is used. Therefore, previously
obtained  values of , for example, $H_0$
from very distant objects with large $z$  based on the standard cosmology
have to be reevaluated in the modified Friedman cosmology. This will be
discussed elsewhere.

\acknowledgments

The author would like to thank D.~Dominici, and the collaborators of
the paper cited in Ref.\cite{seven}, A.~Sinibaldi and J.~Song
for many important discussions and
help. He would also like to express his gratitude to A.~Bottino, R.~Casalbuoni,
C.~Giunti, M.~Im, and L.~Lusanna for helpful discussions. This work was
initiated by the
author's conversation with O.~Bertolami who pointed out some earlier work, to
whom the author is grateful. This work was supported, in part, by the National
Science Foundation, U.S.A.

\end{document}